\documentclass[twocolumn,american,aps,pra,superscriptaddress,floatfix,showpacs]{revtex4}
\usepackage{amsmath}
\usepackage{amssymb}
\usepackage{amsthm}
\usepackage{ulem}
\usepackage{bm}
\usepackage{subfigure}
\usepackage[hypertex]{hyperref}
\usepackage{graphicx}
\usepackage{MnSymbol}
\usepackage{xcolor}

\begin{document}

\title{Fidelity threshold of the surface code beyond single-qubit
  error models}

\author{Pejman Jouzdani}

\affiliation{Department of Physics, University of Central Florida,
  Orlando, Florida 32816, USA}

\author{E. Novais}

\affiliation{Centro de Ci\^encias Naturais e Humanas, Universidade
  Federal do ABC, Santo Andr\'e, SP, Brazil}

\author{I. S. Tupitsyn}

\affiliation{Physics Department, University of Massachusetts, Amherst,
  MA 01003, USA}

\affiliation{Pacific Institute of Theoretical Physics, University of
  British Columbia, Vancouver, BC V6T1Z1, Canada}

\author{Eduardo R. Mucciolo}

\affiliation{Department of Physics, University of Central Florida,
  Orlando, Florida 32816, USA}

\date{\today}


\begin{abstract}
The surface code is a promising alternative for implementing
fault-tolerant, large-scale quantum information processing. Its high
threshold for single-qubit errors under stochastic noise is one of its
most attractive features. We develop an exact formulation for the
fidelity of the surface code that allows us to probe much further on
this promise of strong protection. This formulation goes beyond the
stochastic single-qubit error model approximation and can take into
account both correlated errors and inhomogeneities in the coupling
between physical qubits and the environment. For the case of a
bit-flipping environment, we map the complete evolution after one
quantum error correction cycle onto the problem of computing
correlation functions of a two-dimensional Ising model with boundary
fields. Exact results for the fidelity threshold of the surface code
are then obtained for several relevant types of noise. Analytical
predictions for a representative case are confirmed by Monte Carlo
simulations.
\end{abstract}

\maketitle

\section{Introduction}

Quantum error correction (QEC) is one of the most important tools to
reduce the effects of decoherence in quantum systems that process
information. Several different protocols have been developed since QEC
was first introduced \cite{shor}, but particular attention has been
given to stabilizer codes \cite{gottesman}. Among them, the surface
code \cite{bravyi1998,dennis2002} is perhaps the most promising for
large-scale implementations \cite{fowler2012b}. Its main virtues are:
(i) qubits are disposed in a planar array, only requiring local
measurement operations; and (ii) early estimates based on stochastic
error models indicated a very large threshold value, $p_{c}\approx
11\%$ \cite{dennis2002}, for the single-qubit error probability
$p$. For $p<p_{c}$, the probability of successful encoding tends to 1
as the number of physical qubits is increased.

Despite the large theoretical effort that has been devoted to
characterizing the threshold of the surface code \cite{dennis2002,
  raussendorf, wang2011, ghosh2012, fowler, stephens2013}, the true
nature of the transition has been hard to assess due to the large
Hilbert space that the code demands. Some criticism has also been
raised by the use of simplified error models in these studies, since,
for more traditional QEC schemes, error models that take into account
correlations can substantially alter or even remove error thresholds
\cite{klesse2005, novais2007, preskill}. In this paper we make
significant progress on both issues. We consider more general bit-flip
error models with and without disorder and correlations. We derive an
exact mapping of a complete QEC quantum evolution with arbitrary
syndrome onto a two-dimensional Ising model with complex
temperature. Exact results are then obtained for what we call the
``one-cycle threshold''. Our main conclusion is a positive one: a
fidelity threshold exists in most cases, although its value is not
universal, depending on the noise model. For a representative case,
the analytical prediction for the threshold location based on the
mapping is supported by Monte Carlo simulations.

The remaining of the paper is organized as follows. In
Sec. \ref{sec:threshold} we discuss the difference between intrinsic
and effective thresholds, which is crucial for the understanding of
our results. The next two sections are mainly a review:
Sec. \ref{sec:surfacecode} contains a concise description of the
surface code and Sec. \ref{sec:evolution} describes the code's
evolution, syndrome extraction, and error correction within one cycle
in very general terms. The description of our work begins in
Sec. \ref{sec:fidelity}, where some basic assumptions and definitions
are provided and a suitable expression for the fidelity is
presented. This is followed by a discussion in Sec. \ref{sec:decoding}
of decoding and the thermodynamic limit in the determination of the
threshold. A realistic error model that induces bit-flip errors is
introduced in Sec. \ref{sec:errormodel} and consists of an effective
action involving single-qubit and two-qubit interaction terms. Using
this error model and considering the full quantum evolution of the
physical qubits, in Sec. \ref{sec:mapping} we map the fidelity
calculation after one QEC cycle onto the evaluation of correlation
functions of a two-dimensional Ising model. In Sec. \ref{sec:cases} we
discuss several scenarios based on that mapping, including cases with
homogeneous and inhomogeneous couplings. In Sec. \ref{sec:numerics} we
present the result of Monte Carlo simulations of the fidelity
threshold and confirm the analytical prediction based on the mapping
for the homogeneous coupling case. Conclusions are provided in
Sec. \ref{sec:conclusions}

\section{Intrinsic and effective thresholds}
\label{sec:threshold}

Two QEC strategies can be used for any stabilizer code. In the
so-called active QEC, stabilizer operators are measured and, based on
their syndromes, a recovery operation is chosen and implemented. In
passive QEC, the physical qubits are subjected to a Hamiltonian that
enforces an energy gap between the code word subspace and the rest of
the Hilbert space of the physical qubits. Typically, the Hamiltonian
consists of a sum over all stabilizer operators multiplied by a
negative constant. Protection in this case requires neither
measurements nor recovery operations. While the surface code is an
example of active QEC, the toric code \cite{kitaev} is its passive
counterpart.

Now, consider adding to the toric code a perturbation (e.g., the
environmental noise) that acts directly on the physical qubits and
competes with the code's intrinsic Hamiltonian. Several authors have
shown that beyond a certain critical value of the perturbation's
coupling constant, the toric Hamiltonian is no longer capable of
protecting the code word subspace; topological order is completely
lost and so is the spectral gap separating the ground state subspace
from the excited states \cite{trebst2007,tupitsyn2010,terhalRMP}. In
this context, it is clear that the critical value of the coupling
constant provides an intrinsic threshold: any perturbation larger than
the threshold renders the code completely ineffective even when the
code distance is increased. It is natural to assume that different
types of perturbations will have distinct thresholds.

In this paper we extend the concept of intrinsic thresholds to active
QEC. While for passive QEC the intrinsic threshold due to
environmental noise reveals itself as a clear-cut quantum phase
transition, in active QEC the situation is a more subtle because the
outcome of the QEC cycle depends on the syndromes, their decoding, and
the recovery operation. Thus, it is natural to look for the intrinsic
threshold in the most favorable situation, one that is not affected by
a particular decoding strategy for codes. This happens in the case of
a nonerror syndrome, when no recovery operation is recommended.

For any other syndrome, the threshold must be less favorable because
there can be a certain amount of uncertainty as to which recovery
operation is more effective.

Within this approach, we distinguish two kinds of thresholds:

\begin{itemize}
\item An intrinsic one, which depends only on the interaction between
  the physical qubits and the environment and is independent of any
  decoding procedure.
\item An effective one, which depends on the interaction between the
  physical qubits and the environment and on the decoding procedure.
\end{itemize}

This distinction is valid for any stabilizer code. The effective
threshold is always equal or smaller than the intrinsic threshold. The
effective threshold can always be increased by improving the decoding
procedure until it reaches the intrinsic value. Thus, perfect decoding
makes the effective threshold equal to the intrinsic one in the limit
when the code distance goes to infinity.

In the context of the surface code, the existence of an intrinsic
threshold has been demonstrated by associating the threshold to a
phase transition of a classical statistical spin model with quenched
disorder \cite{dennis2002,terhalRMP}. The proof assumes that physical
qubits are subjected to independent depolarizing noise sources
uncorrelated in time. In this paper we show that an intrinsic
threshold also exists for a noise source where spatial correlations
among physical qubits are induced by their interaction with a common
environment. The threshold in this case is associated to a
finite-temperature phase transition of a two-dimensional Ising model,
with the coupling constant between physical qubits and the environment
playing the role of the inverse temperature of the model.

\section{The surface code and the stabilizer formalism}
\label{sec:surfacecode}

In a QEC stabilizer protocol, information is encoded into a much
larger Hilbert space than the minimum space physically
required. Different sectors of this large Hilbert space are labeled by
different values of observables associated to operators known as
stabilizer. A judicial choice for the stabilizers can then be used to
diagnose the most common type of error for a given quantum
evolution. Based on the outcomes (syndromes) of measurements of
stabilizer operators, a forceful return to the logical Hilbert space
is performed.

\begin{figure}
\centering
\includegraphics[width=6cm]{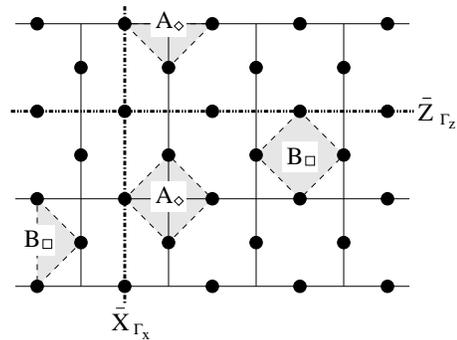}
\caption{Schematic representation of the surface code. The circles are
  physical qubits. The gray squares represent plaquettes ($B_\square$)
  and stars ($A_\diamondsuit$). The product of single-qubit operators
  along the dashed-dotted lines $\Gamma_X$ (vertical) and $\Gamma_Z$
  (horizontal) define the logical operators $\bar{X}_{\Gamma_X}$ and
  $\bar{Z}_{\Gamma_Z}$, respectively.}
\label{fig:surfacecode} 
\end{figure}

The surface code consists of a two-dimensional array of qubits placed
on the edges of a square lattice, see
Fig. \ref{fig:surfacecode}. These physical qubits can be implemented
with Josephson junctions \cite{superconductor}, cold atoms
\cite{coldatoms}, trapped ions \cite{trappedions}, Rydberg atoms
\cite{rydberg}, or semiconductor quantum dots
\cite{semiconductor}. The stabilizers of the code are the plaquette
operators 
\begin{equation}
B_{\square} = \prod_{{\bf i}\in\square} \sigma_{{\bf i}}^{z}
\end{equation}
and the star operators
\begin{equation}
A_{\diamondsuit} = \prod_{{\bf i}\in\diamondsuit} \sigma_{{\bf
    i}}^{x},
\end{equation}
where $\sigma_{i}^{x,z}$ are the Pauli operators $x$ and $z$ of qubit
$i$. The logical operations are defined as strings of physical qubit
operations:
\begin{equation}
\bar{X}_{\Gamma_{X}} = \prod_{{\bf i}\in\Gamma_{X}} \sigma_{{\bf
    i}}^{x}
\end{equation}
and
\begin{equation}
\bar{Z}_{\Gamma_{Z}} = \prod_{{\bf i}\in\Gamma_{Z}}\sigma_{{\bf
    i}}^{z},
\end{equation}
where $\Gamma_{Z}$ is a path that cuts through the lattice from left
to right and $\Gamma_{X}$ is a path that goes from top to bottom
\cite{obs1}. Finally, the codewords can be written as
\begin{equation}
\label{eq:codeword0}
\left| \bar{\uparrow} \right\rangle = G \left| F_{z} \right\rangle
\end{equation}
and
\begin{equation}
\label{eq:codeword1}
\left| \bar{\downarrow} \right\rangle = \bar{X}_{\Gamma_{X}} G \left|
F_{z} \right\rangle,
\end{equation}
where
\begin{equation}
\label{eq:G}
G = \frac{1}{\sqrt{2^{N_{\diamondsuit}}}} \prod_{\diamondsuit} \left(
1 + A_{\diamondsuit} \right)
\end{equation}
and 
\begin{equation}
\label{eq:ferro}
\left|F_{z}\right\rangle = \left|\uparrow\right\rangle_{\bf 1} \otimes
\cdots \otimes \left|\uparrow\right\rangle_{\bf N}
\end{equation}
is the ferromagnet state in the $z$ component of the physical
qubits. Here, $N_{\diamondsuit}$ denotes the number of star operators
on the lattice and $N$ is the total number of physical qubits. The
product $\bar{X}_{\Gamma_{X}}G$ is independent of the particular
choice of $\Gamma_{X}$ and uniquely defines $\left| \bar{\downarrow}
\right\rangle$ (thus, hereafter we will drop the path subscript in the
logical operators). The logical space is a two-dimensional Hilbert
space where all plaquettes and stars, when measured, return the value
$+1$; by convention, this set is called the zero charge sector. All
other sets define sectors with nonzero charge and are characterized by
the number of $-1$ syndrome values for the plaquette and star
operators.

\section{Quantum evolution, syndrome, and error correction}
\label{sec:evolution}

To make the discussion more concise, and without loss of generality
for the case of bit-flip errors, let us assume that the system is
initially prepared in the logical state $\left| \bar{\uparrow}
\right\rangle$ and is not entangled with the environment
$|e\rangle$,
\begin{equation}
|\Psi(0)\rangle = \left| \bar{\uparrow},e \right\rangle.
\end{equation}
The logical qubit and the environment evolve under a unitary evolution
operator $U(\Delta)$ for a time $\Delta$,
\begin{equation}
|\Psi(\Delta)\rangle = U(\Delta)\, |\Psi(0)\rangle.
\end{equation}

Since we are assuming only bit-flip errors, stars will remain always
with eigenvalue $1$ under this evolution. On the other hand,
plaquettes may have eigenvalue $\pm 1$. Let us call $\left\{p\right\}$
a set of plaquettes that return a nontrivial syndrome, thus indicating
an error. When the stabilizers are measured, the system's state vector
is projected by the operator
\begin{equation}
{\cal P}_{\{p\} } =\frac{1}{2^{N_{\square}}} \prod_{\square^{\prime}}
\left(1+B_{\square^{\prime}}\right)\prod_{p}\left(1-B_{p}\right),
 \end{equation}
where $\square^{\prime}$ is the set of plaquettes with eigenvalue
$+1$. The decoding procedure associates to a syndrome a certain
recovery operation. In the surface code, this corresponds to choosing
a string ${\cal S}_{\left\{ p\right\} }^{x}$ made of a product of
$\sigma_{{\bf i}}^{x}$ operators that connect the plaquettes in
$\left\{ p\right\}$ pairwise or to the boundaries. This recovery
operation results in the unnormalized state vector
\begin{eqnarray}
\label{projected_evolution}
|\Psi_{\rm QEC}\rangle_{\{p\}} & = & {\cal S}_{\left\{ p\right\} }^{x}
{\cal P}_{\{p\} } |\Psi(\Delta) \rangle \nonumber \\ & = & {\cal
  S}_{\left\{ p\right\} }^{x} {\cal P}_{\{p\} } U\left( \Delta \right)
\left| \bar{\uparrow}, e\right\rangle.
\end{eqnarray}
A small but important simplification can be made: notice that
\begin{eqnarray}
{\cal S}_{\left\{ p\right\}}^{x} {\cal P}_{\{p\}} & = &
\frac{1}{2^{N_{\square}}} {\cal S}_{\left\{ p\right\}}^{x}
\prod_{\square^{\prime}} \left( 1+B_{\square^{\prime}} \right)
\prod_{p} \left( 1-B_{p} \right) \nonumber \\ & = &
\frac{1}{2^{N_{\square}}} \prod_{\square} \left( 1+B_{\square} \right)
     {\cal S}_{\left\{ p\right\} }^{x},
\end{eqnarray}
where in the second line the product over plaquettes is
unconstrained. Therefore,
\begin{equation}
\label{eq:PsiQEC}
|\Psi_{\rm QEC}\rangle_{\{p\}} = \frac{1}{2^{N_{\square}}}
\prod_{\square} \left( 1+B_{\square} \right) {\cal S}_{\left\{
  p\right\}}^{x} U(\Delta) \left| \bar{\uparrow}, e\right \rangle.
\end{equation}
The presence of the projector $\prod_{\square} (1+B_{\square})$ on the
right-hand side of Eq. (\ref{eq:PsiQEC}) implies that $|\Psi_{\rm
  QEC}\rangle$ contains only qubit states with all plaquettes
positive, i.e., it can be represented by a superposition of the two
logical states $|\bar{\uparrow}\rangle$ and
$|\bar{\downarrow}\rangle$. The amplitude of each state depends on the
choice of ${\cal S}_{\{p\}}$. A good choice will favor
$|\bar{\uparrow}\rangle$.

Even though there is a large number of strings ${\cal
  S}_{\left\{p\right\} }^{x}$ compatible with the syndrome
$\left\{p\right\}$, they can be sorted into two distinct classes
related by the logical operator $\bar{X}$. The product ${\cal
  S}_{\left\{p\right\}}^{x}G$, implicit in the evolution of the
logical qubit in Eq. (\ref{eq:PsiQEC}), generates all possible strings
within a class. Therefore, the particular choice of ${\cal
  S}_{\left\{p\right\}}^{x}$ to represent a class is irrelevant to the
calculation of the fidelity of the code. After choosing a string
${\cal S}_{\left\{ p\right\} }^{x}$, we can assign $\bar{X} {\cal
  S}_{\left\{p\right\}}^{x}$ to represent the elements of the other
class.

Thus, if the recovery operation ${\cal S}_{\left\{p\right\}}^{x}$
brings the logical qubit state back to $|\bar{\uparrow}\rangle$, its
counterpart $\bar{X} {\cal S}_{\left\{p\right\}}^{x}$ takes it to the
state $|\bar{\downarrow}\rangle$. We can then write the (unnormalized)
state vector at the end of the QEC cycle as
\begin{equation}
\label{eq:Psidecomp}
|\Psi_{\rm QEC}\rangle_{\{p\}} = | \bar{\uparrow} \rangle \langle
\bar{\uparrow} | {\cal S}_{\left\{ p\right\}}^{x} U(\Delta) |
\bar{\uparrow}, e \rangle + | \bar{\downarrow} \rangle \langle
\bar{\uparrow} | \bar{X} {\cal S}_{\left\{ p\right\} }^{x} U(\Delta) |
\bar{\uparrow}, e \rangle.
\end{equation}
[Notice the partial contraction in the expectation values appearing on
  the right-hand side of Eq. (\ref{eq:Psidecomp}); environmental
  degrees of freedom remain non contracted.] In the case of a bad
recovery operation, the roles of ${\cal S}_{\left\{p\right\}}^{x}$ and
$\bar{X} {\cal S}_{\left\{p\right\}}^{x}$ are swapped. The fact that
one cannot be completely sure of the efficacy of the recovery
operation is the reason why the effective threshold is always equal or
smaller than the intrinsic threshold.

\section{Environment resetting and the one-cycle fidelity}
\label{sec:fidelity}

The fidelity of the code after one QEC cycle with syndrome set
${\{p\}}$ is defined as
\begin{equation}
{\cal F}_{\{p\}} \equiv \frac{\left| \langle \Psi(0)| \Psi_{\rm
    QEC}\rangle_{\{p\}} \right|^2} {|| \Psi_{\rm QEC} ||^2}.
\end{equation}
Maintaining maximum fidelity at the end of the QEC cycle implies ${\cal
  F}_{\{p\}}=1$ while complete loss of fidelity results in ${\cal
  F}_{\{p\}}=\frac{1}{2}$.

In the evaluation of the fidelity, an important simplifying hypothesis
can be used. It is physically reasonable to assume that the
environment's excitations can be suppressed by some ``cooling''
mechanism (e.g., lowering the temperature, applying a polarizing
field, etc). This assumption was previously discussed in
Refs. \cite{novais2013,jouzdani2013}. Assuming this resetting of the
environment at the end of the QEC cycle, we can rewrite
Eq. (\ref{eq:Psidecomp}) as
\begin{equation}
|\Psi_{\rm QEC}\rangle_{\{p\}} = {\cal A}_{\{p\}}
|\bar{\uparrow},e\rangle + {\cal B}_{\{p\}}
|\bar{\downarrow},e\rangle,
\end{equation}
where we have introduced the amplitudes
\begin{equation}
\label{eq:A}
{\cal A}_{\left\{ p\right\} } = \left\langle
\bar{\uparrow},e\right|{\cal S}_{\left\{ p\right\}
}^{x}U\left(\Delta\right)\left|\bar{\uparrow},e\right\rangle
\end{equation}
and
\begin{equation}
\label{eq:B}
{\cal B}_{\left\{ p\right\} } = \left\langle
\bar{\uparrow},e\right|\bar{X}{\cal S}_{\left\{ p\right\}
}^{x}U\left(\Delta\right)\left|\bar{\uparrow},e\right\rangle ,
\end{equation}
for each syndrome outcome characterized by the set $\{p\}$. As a
result, after some simple manipulations \cite{novais2013}, we can
write
\begin{equation}
\label{eq:fidelity}
{\cal F}_{\left\{ p\right\} } = \frac{\left| {\cal A}_{\left\{
    p\right\} } \right|^{2}} {\left|{\cal A}_{\left\{ p\right\} }
  \right|^{2} + \left| {\cal B}_{\left\{ p\right\} } \right|^{2}}.
\end{equation}

When the couping between the physical qubits and the environment is
sufficiently weak, one can expand the evolution operator $U(\Delta)$
in terms of strings of operators $\sigma_{\bf i}^x$ of increasing
length. Each string represents a certain number of bit-flip events,
with that number defining the length of the string. Looking at
Eq. (\ref{eq:A}), one recognizes that the shortest string in the
expansion that gives a nonzero contribution to ${\cal A}_{\{p\}}$
corresponds exactly to the string of $\sigma_{\bf i}^x$ operators in
${\cal S}_{\{p\}}$. Therefore, the order of the leading contribution
to ${\cal A}_{\{p\}}$ is equal to the smallest possible length of
${\cal S}_{\{p\}}$. Similarly, the leading contribution to ${\cal
  B}_{\{p\}}$ has an order equal to the smallest possible length of
$\bar{X}\, {\cal S}_{\{p\}}$. Therefore, in the weak-coupling limit,
whenever ${\cal S}_{\{p\}}$ is shorther than $\bar{X}\, {\cal
  S}_{\{p\}}$, one expects $|{\cal A}_{\{p\}}|>|{\cal B}_{\{p\}}|$. In
the strong coupling limit, on the other hand, any perturbative
expansion fails; in those circumstances, we expect ${\cal A}_{\{p\}}$
and ${\cal B}_{\{p\}}$ to have similar magnitudes on a finite lattice.

The dependence of ${\cal A}_{\{p\}}$ and ${\cal B}_{\{p\}}$ on the
lenghts of ${\cal S}_{\{p\}}$ and $\bar{X}\, {\cal S}_{\{p\}}$,
respectively, give us a hint to the appropriate decoding procedure and
thermodynamic limit we need to adopt in order to define a fidelity
threshold.

\section{Decoding and the thermodynamic limit}
\label{sec:decoding}

The expression for the fidelity given in Eq. (\ref{eq:fidelity}) is
only valid under the assumption that the decoding of the syndromes is
flawless, such that $\left| {\cal A}_{\{p\}} \right| > \left| {\cal
  B}_{\{p\}} \right|$. In practice, finding a recovery operation
${\cal S}_{\{p\}}^{x}$ that fulfills this inequality for any given
syndrome set $\{p\}$ is a difficult task. A rich literature exists on
decoding algorithms for the surface code; see, for instance,
\cite{dennis2002,poulin,bravyi,ferris} and references therein, where a
number of strategies have been proposed. Here, we do not attempt to
improve on the existing strategies. We offer instead a prescription
where the inequality is always satisfied in the limit of infinite code
distance (i.e., infinite lattice sizes). As we argue below, when the
inequality holds, the fidelity threshold obtained from from
Eq. (\ref{eq:fidelity}) is equal to the intrinsic threshold. However,
since it is not possible to guarantee that the decoding is flawless in
practice, at times we will have $\left| {\cal A}_{\{p\}} \right| <
\left| {\cal B}_{\{p\}} \right|$. Therefore, an effective fidelity
threshold that takes into account the possibility of flawed recovery
operations should always be smaller than the intrinsic one.

Our prescription for enforcing $\left| {\cal A}_{\{p\} } \right| >
\left| {\cal B}_{\{p\}} \right|$ begins by recognizing that the
amplitudes ${\cal A}_{\{ p\} }$ and ${\cal B}_{\{ p\}}$ are functions
of the lattice size used to encode the logical qubit. Therefore, it is
important to define what we call the {\it thermodynamic limit}. Let us
start with a fixed and finite set $\{p\}$ containing an even number of
plaquettes and assume that the limit is taken by constructing a
sequence of lattices of increasing size. Consider the smallest lattice
that can contain the set $\{p\}$ as the initial element of the
sequence. The next element in the sequence is constructed from the
previous one by adding rows and columns to all four boundaries of the
lattice. As a consequence, \emph{the distance from any plaquette in
  $\{p\}$ to the boundaries increases with increasing lattice
  sizes}. We now choose a string ${\cal S}_{\{p\}}^{x}$ made of a
product of $\sigma_{{\bf i}}^{x}$ operators that connect pairwise the
plaquettes in $\{p\}$ without reaching the boundaries. By this choice,
the string operator $\bar{X}{\cal S}_{\{p\}}$ used in the computation
of ${\cal B}_{\{p\}}$ always reaches the boundaries while ${\cal
  S}_{\{p\}}^{x}$ does not. As the lattice grows, the length of
$\bar{X}{\cal S}_{\{p\}}$ surpasses that of ${\cal S}_{\{p\}}^{x}$. As
described in Sec. \ref{sec:fidelity}, in the weak coupling limit,
${\cal B}_{\{p\}}$ is strongly suppressed in comparison to ${\cal
  A}_{\{p\}}$. Thus our choice of ${\cal S}_{\{p\}}^{x}$ ensures that,
in the limit of infinite lattice size, the inequality $\left| {\cal
  A}_{\{p\}} \right| > \left| {\cal B}_{\{p\}} \right|$ is satisfied.

The prescription needs to be slightly modified when $\{p\}$ contains
an odd number of plaquettes. After connecting all but one plaquette
with strings pairwise, we connect the unpaired plaquette (presumably
the most remote one) by a string to the closest boundary. We then keep
that boundary fixed and construct the sequence of lattices by adding
columns to the left and right boundaries but adding rows only to the
opposite boundary. Thus, the one string in ${\cal S}_{\{p\}}^{x}$
reaching a boundary will keep its length fixed, while the counterpart
of that string in $\bar{X}{\cal S}_{\{p\}}$ will reach the opposite
boundary with an increasing length. This guarantees that $\left| {\cal
  A}_{\{p\}} \right| > \left| {\cal B}_{\{p\}} \right|$ is also
satisfied.

With this definition of the thermodynamical limit at hand, we define
the one-cycle threshold for the surface code as the largest value of
coupling between the physical qubits and the environment such that, in
the {\it thermodynamic limit}, ${\cal F}_{\{p\}} \to 1$ {\it for any}
finite set $\{p\}$. In particular, the intrinsic threshold corresponds
to the non-error syndrome, namely, when $\{p\}$ is an empty set. A
proof that an intrinsic threshold always exists for the error model
defined in Sec. \ref{sec:errormodel} is given in Appendix
\ref{sec:appendixA}. Our prescription for the thermodynamic limit also
allows us to show in Sec. \ref{sec:mapping} that, for an error model
with only nearest-neighbor qubit interactions, the fidelity threshold
derived from Eq. (\ref{eq:fidelity}) is independent of the syndrome
set $\{p\}$ and thus equal to the intrinsic threshold.

In our prescription for the thermodynamic limit we first fix the
syndrome set $\{p\}$ and then take the lattice size to infinity. Thus,
the density of plaquettes with eigenvalues $-1$ tends to zero. This
procedure is quite adequate for our goal of establishing the intrinsic
threshold irrespective of the syndrome $\{p\}$, as we argue in
Sec. \ref{sec:cases}.

An alternative definition of the thermodynamic limit, commonly used in
numerical investigations using stochastic error models, would keep the
density of plaquettes with eigenvalues $-1$ fixed as the lattice size
is increased. Therefore, the set $\{p\}$ would be different and larger
for each new lattice size and a recalculation of the amplitudes ${\cal
  A}_{\{p\}}$ and ${\cal B}_{\{p\}}$ would be required at each new
step. Thus, in addition to being much harder to be analyzed, such a
prescription does not guarantee that the inequality ${\cal
  A}_{\{p\}}|>|{\cal B}_{\{p\}}|$ is satisfied in the thermodynamic
limit. Therefore, it is not suitable for determining a fidelity
threshold from Eq. (\ref{eq:fidelity}).

\section{The error model}
\label{sec:errormodel}

In realistic implementations, the physical qubits interact with a
variety of environmental degrees of freedom \cite{stamp}. For
instance, frequently one cannot neglect the interaction of qubits with
bosonic environments \cite{novais2007}. These can come directly from
phonons and electromagnetic fluctuations generated by electronic
components, or indirectly from interactions with spin or charge
impurities. It is also possible that imperfections in qubit design
cause spurious coupling between single-qubit states. Finally, it is
also possible that qubits couple to a spin (or pseudo-spin) bath. For
these cases and others, the effect of a time evolution under the
influence of the environment can be recast as an effective action for
the physical qubits by integrating out the environmental degrees of
freedom. Thus, at the end of a QEC cycle, an effective evolution
operator of the form 
\begin{equation}
\label{eq:Ueff}
U_{\rm eff} = \langle e| U(\Delta) | e\rangle = e^{-H_{\rm eff}}
\end{equation}
can be derived and employed in the calculation of matrix elements and
probability amplitudes involving physical qubits. The exact form of
$H_{\rm eff}$ depends on the particular type of interaction and the
nature of the physical qubits and the environment. Here, we consider
the two-term expression
\begin{equation}
\label{eq:Heff}
H_{{\rm eff}} \left( \{\sigma_{\bf i}^x\} \right) = \sum_{{\bf i}}
h_{{\bf i}}\, \sigma_{{\bf i}}^{x} + \sum_{{\bf i} \neq {\bf j}}
J_{{\bf i},{\bf j}}\, \sigma_{{\bf i}}^{x} \sigma_{{\bf j}}^{x}.
\end{equation}
The parameters $h_{{\bf i}}$ and $J_{{\bf i}{\bf j}}$ incorporate
environmental fields and qubit-qubit interactions, respectively, and
can be either real or imaginary numbers. Their magnitudes set the
strength of the coupling between physical qubits and the
environment. This form is exact for qubits coupled linearly to free
bosonic baths and local fields \cite{jouzdani2013}. We note that
$J_{{\bf i}{\bf j}}$ can also represent direct interactions between
qubits that are not environment mediated.

\section{Mapping onto an unconstrained Ising model}
\label{sec:mapping}

Let us write expressions for the amplitudes ${\cal A}_{\{p\}}$ and
${\cal B}_{\{p\}}$ as sums over configurations of the variables
$\{\sigma_{\bf i}^x\}$. We begin by replacing each state
$\left|\uparrow\right\rangle_{\bf i}$ in Eq. (\ref{eq:ferro}) by
$\frac{1}{\sqrt{2}} \left( |+\rangle_{\bf i} + |-\rangle_{\bf i}
\right)$, where $\hat{\sigma}_{\bf i}^x |\pm \rangle_{\bf i} = \pm
|\pm \rangle_{\bf i}$. Introducing the notation $\hat{\sigma}_{\bf
  i}^x |\sigma_{\bf i}\rangle = \sigma_{\bf i} |\sigma_{\bf
  i}\rangle$, we have
\begin{equation}
\label{eq:codeword1new}
| \bar{\uparrow} \rangle = \sum_{\sigma}^{}{}^\prime\, \left|\sigma
\right\rangle
\end{equation}
up to a normalization factor. Here $\sigma$ stands for $(\sigma_{\bf
  1}, \dots, \sigma_{\bf N})$. The sum in Eq. (\ref{eq:codeword1new})
is restricted to the configurations $\sigma$ that satisfy the
constraint $A_\diamondsuit=+1$ for all stars (i.e., vertices) in the
lattice. We substitute Eq. (\ref{eq:codeword1new}) in the definitions
of the amplitudes ${\cal A}_{\{p\}}$ and ${\cal B}_{\{p\}}$
[Eqs. (\ref{eq:A}) and (\ref{eq:B})] and use Eq. (\ref{eq:Ueff}) to
arrive at
\begin{equation}
\label{eq:amplitudeA}
{\cal A}_{\{p\} } = \sum_{\sigma}^{}{}^\prime\, {\cal S}_{\{p\}}^{x}\,
e^{-H_{\rm eff} \left( \sigma \right)}
\end{equation}
and 
\begin{equation}
\label{eq:amplitudeB}
{\cal B}_{\{p\} } = \sum_{\sigma}^{}{}^\prime \bar{X}\, {\cal
  S}_{\{p\}}^{x}\, e^{-H_{\rm eff} \left( \sigma \right)}.
\end{equation}
The operator $S_{\{p\}}^x$ now represents a string of variables
$\sigma_{{\bf i}}$ compatible with the syndrome represented by the set
of plaquettes $\{p\}$.

The sums in Eqs. (\ref{eq:amplitudeA}) and (\ref{eq:amplitudeB}) are
very difficult to evaluate (see Ref. \cite{novais2013}). Below, we
provide an exact solution to this problem for a particular but
significant case.

\begin{figure}
\centering
\includegraphics[width=6cm]{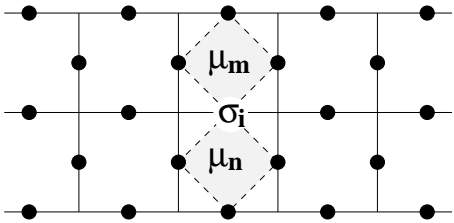}
\caption{Qubit variable $\sigma_{\bf i}$ and corresponding plaquette
  variables $\mu_{\bf m}$ and $\mu_{\bf n}$.}
\label{fig:massfield}
\end{figure}

\subsection{Nearest-neighbor correlations}

\begin{figure}
\centering
\includegraphics[width=6cm]{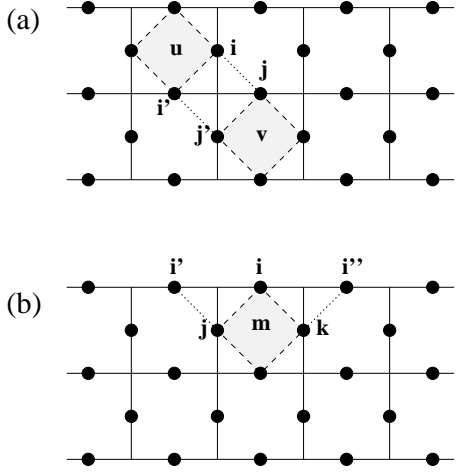}
\caption{(a) Links and plaquettes involved in Eq. (\ref{eq:mapJ}). (b)
  Links and plaquette involved in Eq. (\ref{eq:hJboundary}).}
\label{fig:links}
\end{figure}

Let us consider the case where $J_{{\bf i}{\bf j}}$ describes only
nearest-neighbor interactions. To overcome the restriction in the sums
in Eqs. (\ref{eq:amplitudeA}) and (\ref{eq:amplitudeB}), we introduce
plaquette variables $\{\mu_{\bf m}\}$, with $\mu_{\bf m}=\pm 1$, such
that $\sigma_{{\bf i}} = \mu_{{\bf m}}\, \mu_{{\bf n}}$ (see
Fig. \ref{fig:massfield}). The subscript ${\bf i}$ refers to the
physical qubit ${\bf i}$ while ${\bf m}$ and ${\bf n}$ indicate the
plaquettes sharing the link ${\bf i}$. Even though the variables
$\{\mu_{\bf m}\}$ can be positive or negative, they automatically
satisfies the constraint of positive star eigenvalues. This
parameterization is well known in the lattice gauge field literature
\cite{polyakov}. Thus, starting from the error model defined in
Eq. (\ref{eq:Heff}) and introducing these new variables, it is
straightforward to show that, for nearest-neighbor qubits ${\bf i}$
and ${\bf j}$ in the bulk (i.e., not at the top or bottom edges),
\begin{equation}
h_{\bf i}\, \sigma_{\bf i}^x \longrightarrow h_{\bf i}\, \mu_{\bf m}
\mu_{\bf n}
\end{equation}
and
\begin{equation}
\label{eq:mapJ}
J_{{\bf i}{\bf j}}\, \sigma^x_{\bf i}\, \sigma^x_{\bf j}
\longrightarrow J_{{\bf i}{\bf j}}\, \mu_{\bf u}\, \mu_{\bf v},
\end{equation}
where the plaquettes ${\bf u}$ and ${\bf v}$ are next-to-nearest
neighbors containing the links ${\bf i}$ and ${\bf j}$, respectively
[see Fig. \ref{fig:links}(a)]. Notice that the same product $\mu_{\bf
  u}\, \mu_{\bf v}$ appears again when we consider the contribution
from the other pair of nearest-neighbor links ${\bf i}^\prime$ and
${\bf j}^\prime$ belonging to these plaquettes. Thus, we can define
$\tilde{h}_{{\bf m}{\bf n}} = h_{\bf i}$ such that
\begin{equation}
\label{eq:hbulk}
\sum_{{\bf i}\in{\rm bulk}} h_{\bf i}\, \sigma^x_{\bf i} =
\sum_{\langle {\bf m},{\bf n}\rangle} \tilde{h}_{{\bf m}{\bf n}}\,
\mu_{\bf m}\, \mu_{\bf n},
\end{equation}
as well as $\tilde{J}_{{\bf u}{\bf v}} = J_{{\bf i}{\bf j}} + J_{{\bf
    i}^\prime{\bf j}^\prime}$ such that
\begin{equation}
\label{eq:Jbulk}
\sum_{\langle{\bf i},{\bf j}\rangle\in{\rm bulk}} J_{{\bf i}{\bf j}}\,
\sigma^x_{\bf i}\, \sigma^x_{\bf j} = \sum_{\llangle {\bf u},{\bf
    v}\rrangle} \tilde{J}_{{\bf u}{\bf v}}\, \mu_{\bf u}\, \mu_{\bf v},
\end{equation}
where $\left\langle {\bf m}, {\bf n} \right\rangle$ are
nearest-neighbor plaquettes and $\left\llangle {\bf u},{\bf v}
\right\rrangle$ are next-to-nearest neighbors. The new parameters are
functions of $h_{{\bf i}}$ and $J_{{\bf i}{\bf j}}$. Notice that for
homogeneous fields and couplings in the bulk, $h_{\bf i}=h$ and
$J_{{\bf i}{\bf j}}=J$, we get $\tilde{h}_{{\bf m}{\bf n}} = \tilde{h}
= h$ and $\tilde{J}_{{\bf u}{\bf v}} = \tilde{J} = 2J$.

A little more work is needed to understand the constraint of positive
stars at the bottom and top boundaries (see
Fig. \ref{fig:surfacecode}). A star at these boundaries is formed by
the product of three qubits; for instance, $A_{1} = \sigma_{1}^{x}\,
\sigma_{2}^{x}\, \sigma_{3}^{x} = 1$. One of these qubits belongs to
the bulk and can be written in terms of plaquette variables,
$\sigma_{2}^{x} = \mu_{1}\, \mu_{3}$. Therefore, we obtain $\mu_{1}
\sigma_{1}^{x}\, \mu_{3}\, \sigma_{3}^{x} = 1$ and, consequently,
$\mu_{1}\, \sigma_{1}^{x} = \mu_{3}\, \sigma_{3}^{x} = \pm 1$. This
can be repeated for all qubits at the boundary. As a result, the
product of any pair $\mu_{{\bf m}}\, \sigma_{{\bf i}}^{x}$ at the
boundaries is a constant $\pm 1$ (here ${\bf m}$ denotes the plaquette
where the boundary link ${\bf i}$ is located.) Thus, for sites at the
top $t$ and bottom $b$ boundaries, we can write
\begin{equation}
h_{\bf i}\, \sigma_{\bf i} \longrightarrow \alpha_{t(b)}\, h_{\bf i}\,
\mu_{\bf m}
\end{equation}
and
\begin{equation}
J_{{\bf i}{\bf j}}\, \sigma_{\bf i}\, \sigma_{\bf j} \longrightarrow
\alpha_{t(b)}\, J_{{\bf i}{\bf j}}\, \mu_{\bf n},
\end{equation}
where $\alpha_{t(b)}=\pm 1$ and ${\bf n}$ is the nearest-neighbor
plaquette to ${\bf m}$ that contains the link ${\bf j}$. Similarly to
the bulk case, pairwise interactions between a qubit at the boundary
and another in the bulk contribute twice to terms containing solely
one plaquette variable. As a result,
\begin{equation}
\label{eq:hJboundary}
\sum_{{\bf i}\in t(b)} h_{\bf i}\, \sigma^x_{\bf i} +
\sum_{\langle{\bf i},{\bf j}\rangle, {\bf i}\in t(b)} J_{{\bf i}{\bf
    j}}\, \sigma_{\bf i}^x\, \sigma_{\bf j}^x = \alpha_{t(b)}
\sum_{{\bf m}\in t(b)} \tilde{h}_{\bf m}\, \mu_{\bf m},
\end{equation}
where $\tilde{h}_{\bf m} = h_{\bf i} + J_{{\bf i}^\prime{\bf j}} +
J_{{\bf i}^{\prime\prime}{\bf k}}$. Here, ${\bf i}^\prime$ and ${\bf
  i}^{\prime\prime}$ are the next-to-nearest neighboring links to
${\bf i}$ along the edge and ${\bf j}$ and ${\bf k}$ are links
belonging to the same plaquette where ${\bf i}$ is located [see
  Fig. \ref{fig:links}(b)].

Combining Eqs. (\ref{eq:hbulk}), (\ref{eq:Jbulk}), and
(\ref{eq:hJboundary}) we obtain
\begin{equation}
\label{H_eff-2}
H_{{\rm eff}}(\{\mu_{\bf m}\}) = \sum_{\left\langle {\bf m},{\bf
    n}\right\rangle} \tilde{h}_{{\bf m}{\bf n}}\, \mu_{{\bf m}}\,
\mu_{{\bf n}} + \sum_{\left\llangle {\bf u},{\bf v}\right\rrangle}
\tilde{J}_{{\bf u}\, {\bf v}}\,\mu_{{\bf u}}\, \mu_{{\bf v}},
\end{equation}
Thus, we can completely eliminate the qubit variables.

We can now go back to Eqs. (\ref{eq:amplitudeA}) and
(\ref{eq:amplitudeB}) and switch the restricted sums over qubit
variables $\sigma$ to unrestricted sums over plaquette variables
$\{\mu_{\bf m}\}$. There are two distinct situations to consider,
depending on the number of plaquettes in the syndrome.

For syndromes with an even number of plaquettes, the choice of strings
in $S^x_{\{p\}}$ will involve either no link at the top or bottom
boundaries or an even number of links on the same boundary. Therefore,
we can write
\begin{equation}
\label{eq:Sxeven}
{\cal S}^x_{\{p\}\, {\rm even}} \longrightarrow \prod_{{\bf m}\in
  \{p\}} \mu_{\bf m}.
\end{equation}
and
\begin{equation}
\label{eq:Xeven}
\bar{X}\, {\cal S}^x_{\{p\}\ {\rm even}} \longrightarrow \alpha_b\,
\alpha_t\, \prod_{{\bf m}\in \{p\}} \mu_{\bf m},
\end{equation}
since any logic operator $\bar{X}$ will link top to bottom
boundaries.

For syndromes with an odd number of plaquettes, at least one string in
$S^x_{\{p\}}$ will have to end in one of the boundaries. By applying
the operator $\bar{X}$ one generates a string ending on the opposite
boundary. Then,
\begin{equation}
\label{eq:Sxodd}
{\cal S}^x_{\{p\}\ {\rm odd}} \longrightarrow \alpha_{t(b)}\,
\prod_{{\bf m}\in \{p\}} \mu_{\bf m}.
\end{equation}
and
\begin{equation}
\label{eq:Xodd}
\bar{X}\, {\cal S}^x_{\{p\}\ {\rm odd}} \longrightarrow
\alpha_{b(t)}\, \prod_{{\bf m}\in \{p\}} \mu_{\bf m}.
\end{equation}
[If the unpaired plaquette is linked to the top boundary by ${\cal
    S}_{\{p\}}^x$, then one picks the subscript $t$ on the right-hand
  side of Eq. (\ref{eq:Sxodd}). The opposite goes for
  Eq. (\ref{eq:Xodd}).]

To obtain new expressions for the amplitudes ${\cal A}_{\{p\}}$ and
${\cal B}_{\{p\}}$ in terms of the variables $\{\mu_{\bf m}\}$,
$\alpha_b$, and $\alpha_t$, one inserts Eqs. (\ref{eq:Sxeven}) and
(\ref{eq:Xeven}) or Eqs. (\ref{eq:Sxodd}) and (\ref{eq:Xodd}) into
Eqs. (\ref{eq:amplitudeA}) and (\ref{eq:amplitudeB}) and sum over all
configurations of the variables $\alpha_t$, $\alpha_b$, and
$\{\mu_{\bf m}\}$. In order to simplify the result, we introduce the
(unnormalized) correlation function
\begin{equation}
\label{eq:Cp}
C_{\{p\}} \left( \alpha_b, \alpha_t \right) = \sum_{\left\{ \mu_{\bf
    m} \right\}} \left( \prod_{{\bf k} \in \left\{ p \right\}}
\mu_{\bf k}\right) \, e^{- \tilde{H}(\{\mu_{\bf
    m}\};\alpha_t,\alpha_b)},
\end{equation}
where
\begin{equation}
\label{eq:Hfull}
\tilde{H}(\{\mu_{\bf m}\};\alpha_t,\alpha_b) = H_{\rm eff}(\{\mu_{\bf
  m}\}) + H_{\rm bound}(\{\mu_{\bf m}\};\alpha_t,\alpha_b), 
\end{equation}
and
\begin{equation}
\label{eq:Hedge}
H_{\rm bound}(\{\mu_{\bf m}\};\alpha_t,\alpha_b) = \alpha_t
\sum_{{\bf t}} \tilde{h}_{\bf t}\, \mu_{{\bf t}} + \alpha_b \sum_{{\bf
    b}} \tilde{h}_{\bf b}\, \mu_{{\bf b}}.
\end{equation}
Notice that since $H_{\rm eff}$ contains only two-body interaction
terms, the following symmetry relation is satisfied:
\begin{equation}
\label{eq:TRS}
C_{\{p\}} (-\alpha_t,-\alpha_b) = (-1)^{N_p}\, C_{\{p\}} (\alpha_t,\alpha_b),
\end{equation}
where $N_p$ is the number of plaquettes in $\{p\}$. Using
Eqs. (\ref{eq:Sxeven}), (\ref{eq:Xeven}), (\ref{eq:Sxodd}), and
(\ref{eq:Xodd}) in conjunction with Eqs. (\ref{eq:Cp}) and
(\ref{eq:TRS}) allows us to obtain concise relations for the
amplitudes ${\cal A}_{\{p\}}$ and ${\cal B}_{\{p\}}$. For $N_p$ even,
we get
\begin{equation}
\label{eq:Ising-gen-Aeven}
{\cal A}_{\{p\}}^{\rm even} = C_{\{p\}}(+,+) + C_{\{p\}}(+,-)
\end{equation}
and
\begin{equation}
\label{eq:Ising-gen-Beven}
{\cal B}_{\{p\}}^{\rm even} = C_{\{p\}}(+,+) - C_{\{p\}}(+,-)
\end{equation}
For $N_p$ odd, there are two situations to consider. When the bottom
is closest boundary to the most remote plaquette in $\{p\}$, we get
\begin{equation}
\label{eq:Ising-gen-Aodd_b}
{\cal A}_{\{p\}}^{\rm odd;\, bottom} = C_{\{p\}}(+,+) - C_{\{p\}}(+,-)
\end{equation}
and
\begin{equation}
\label{eq:Ising-gen-Bodd_b}
{\cal B}_{\{p\}}^{\rm odd;\, bottom} = C_{\{p\}}(+,+) + C_{\{p\}}(+,-)
\end{equation}
while when the closest boundary is the top one,
\begin{equation}
\label{eq:Ising-gen-Aodd_t}
{\cal A}_{\{p\}}^{\rm odd;\, top} = C_{\{p\}}(+,+) + C_{\{p\}}(+,-)
\end{equation}
and
\begin{equation}
\label{eq:Ising-gen-Bodd_t}
{\cal B}_{\{p\}}^{\rm odd;\, top} = C_{\{p\}}(+,+) - C_{\{p\}}(+,-).
\end{equation}
%

\section{The threshold as a phase transition}
\label{sec:cases}

As argued in Sec. \ref{sec:decoding}, our prescription of the
thermodynamic limit guarantees that $|{\cal A}_{\{p\}}| > |{\cal
  B}_{\{p\}}|$ as the lattice size grows and therefore we can use
Eq. (\ref{eq:fidelity}) to evaluate the fidelity. Equations
(\ref{eq:Ising-gen-Aeven}) to (\ref{eq:Ising-gen-Bodd_t}) are exact
expressions, valid for any lattice size and for any finite set
$\{p\}$, provided that the effective interaction between physical
qubits involves only nearest neighbors. Thus, the behavior of ${\cal
  A}_{\{p\}}$ and ${\cal B}_{\{p\}}$ is completely determined by
$\tilde{H}$ and its associated correlation function $C_{\{p\}}$. Since
the boundary-field Ising model defined by $\tilde{H}$ has a
finite-temperature critical point, the fidelity threshold can be
understood as this phase transition point. 

Let us now show that this interpretation is valid. To simplify the
argument, suppose that either $\tilde{h}=0$ and $\tilde{J}\neq0$ or
$\tilde{h}\neq0$ and $\tilde{J}=0$, in which case the Ising model
defined in Eq. (\ref{H_eff-2}) can always be transformed into a
ferromagnetic model (when $\tilde{J}=0$) or two decoupled
ferromagnetic models (when $\tilde{h}=0$) by an appropriate change in
the signs of the the variables $\mu_{\bf m}$ beloging to one of the
sublattices.

\subsection{$N_p$ even}
\label{sec:even}

Consider the case where $N_p$ is even. For temperatures above the
critical value (i.e., for small enough coupling constants), the
$\{\mu_{\bf m}\}$ spin system of Eq. (\ref{H_eff-2}) is in a
disordered (paramagnetic) phase. In the thermodynamic limit, because
both boundaries become infinitely distant from the plaquettes in
$\{p\}$, $C_{\{p\}}$ will not depend on boundary fields $\alpha_b$ and
$\alpha_t$, namely, $C_{\{p\}}(\alpha_t,\alpha_b) \rightarrow
\sum_{\left\{ \mu_{\bf m} \right\}} \left( \prod_{{\bf k} \in \left\{
  p \right\}} \mu_{\bf k}\right) \, e^{- H_{\rm eff}(\{\mu_{\bf m}\})}
= C_{\{p\}}(0,0)$. Notice that even though spatial correlations among
the spins $\{\mu_{\bf k}\}$ in the set $\{p\}$ decay exponentially in
space, they are finite even in the infinite-lattice limit because
their are locked in their positions. Therefore, $|C_{\{p\}}(0,0)|>0$
and ${\cal B}_{\{p\}} \rightarrow 0$, independently of $\{p\}$. As a
result, ${\cal F}_{\{p\}} \rightarrow 1$.

Conversely, for low temperatures (i.e., large enough coupling
constants), the spin system is in an ordered (ferromagnetic) phase. We
can then distinguish two situations: (i) $\alpha_b\, \alpha_t=1$, when
there is an even number of domain walls running parallel to the top
and bottom boundaries; (ii) $\alpha_b\, \alpha_t=-1$, when the number
of domain walls is odd. In both situations the domain walls are rather
costly energetically (the cost scales with $N_{\rm col}$, the number
of columns in the lattice). Thus, for low enough temperatures and
$N_{\rm col} \gg 1$, we can assume that $C_{\{p\}}(\pm,\pm)$ is
governed by spin configurations with no domain wall, whereas
$C_{\{p\}}(\pm,\mp)$ is governed by configurations with just one
domain wall. In addition, since correlations decay exponentially in
space in the ordered phase as well, $\llangle \prod_{{\bf k} \in
  \left\{ p \right\}} \mu_{\bf k} \rrangle \approx \prod_{{\bf k} \in
  \left\{ p \right\}} \llangle \mu_{\bf k} \rrangle$, where $\llangle
\cdots \rrangle = \sum_{\left\{ \mu_{\bf m} \right\}} (\cdots)\, e^{-
  \tilde{H}(\{\mu_{\bf m}\};\alpha_t,\alpha_b)}$, with each $\llangle
\mu_{\bf k} \rrangle=\pm 1$, depending on which side of the domain
wall the site ${\bf k}$ is located. Therefore, since we have an even
number of plaquettes in $\{p\}$, $C_{\{p\}}(\pm,\pm)>0$. For
$C_{\{p\}}(\pm,\mp)$, on the other hand, because the domain wall can
cut across the plaquettes in $\{p\}$ in many different ways with
similar energy costs, this amplitude results from a sum of many
similar terms with alternating signs. $C_{\{p\}}(\pm,\mp)$ is strongly
suppressed with respect to $C_{\{p\}}(\pm,\pm)$ and, as a result,
${\cal A}_{\{p\}} \approx {\cal B}_{\{p\}}$ and ${\cal F}_{\{p\}}
\approx \frac{1}{2}$, independently of the location of the plaquettes
in the set $\{p\}$, as long as they are in finite number.

\subsection{$N_p$ odd}
\label{sec:odd}

Let us now consider cases where $N_p$ is odd and assume that the most
remote plaquette in $\{p\}$ is closer to the top boundary (it is
straightforward to extend the discussion to the opposite situation).

For temperatures higher than the critical one, the spin system is in a
disordered (paramagnetic) phase. In the thermodynamic limit, the
bottom boundary will become infinitely distant to the plaquettes in
$\{p\}$ and $C_{\{p\}}(\alpha_t,\alpha_b)$ will not depend on the
boundary field $\alpha_b$: $C_{\{p\}}(\alpha_t,\alpha_b) \rightarrow
C_{\{p\}}(\alpha_t,0)$. As a result, from
Eqs. (\ref{eq:Ising-gen-Aodd_t}) and (\ref{eq:Ising-gen-Bodd_t}) we
see that ${\cal B}_{\{p\}} \rightarrow 0$, independently of $\{p\}$,
while ${\cal A}_{\{p\}}$ takes a nonzero finite value determined by
the residual spatial correlations between the spin variables
$\{\mu_{\bf k}\}_{{\bf k}\in\{p\}}$. Therefore, ${\cal F}_{\{p\}}
\rightarrow 1$.

The argument for the low-temperature limit follows closely that
developed in Sec. \ref{sec:cases} when the number of plaquettes is
even. In the ordered (ferromagnetic) phase, $C_{\{p\}}(\pm,\pm)$ is
governed by spin configurations with no domain walls. Then,
$C_{\{p\}}(+,+)>0$ and $C_{\{p\}}(-,-)>0$. The correlation functions
$C_{\{p\}}(\pm,\mp)$, on the other hand, are dominated by spin
configurations with a single domain wall running parallel to the top
and bottom boundaries and result from a sum of terms with alternating
signs with roughly the same energy costs. As a result, they are
suppressed in magnitude in comparison to $C_{\{p\}}(\pm,\pm)$. Thus,
using Eqs. (\ref{eq:Ising-gen-Aodd_t}) and
(\ref{eq:Ising-gen-Bodd_t}), we concluded that ${\cal A}_{\{p\}}
\approx {\cal B}_{\{p\}}$ and ${\cal F}_{\{p\}} \approx \frac{1}{2}$.

\subsection{Phase transition}

For both $N_p$ even and odd, the abrupt change in behavior of the
correlation function $C_{\{p\}}$ as the critical temperature is
crossed is what renders the transition from ${\cal F}_{\{p\}} = 1$ to
${\cal F}_{\{p\}} = \frac{1}{2}$ sharp and what defines the location
of the threshold value for the coupling constants.

We stress that this transition is not the same as that originally
discussed in Ref. \cite{dennis2002}, where the error model was a
purely stochastic one with no correlations. By virtue of the
stochastic nature of the errors, their problem mapped onto a spin
glass on the Nishimori line. Instead, the transition we obtain bares
close resemblance to that for the toric code in the presence of a
transverse field \cite{trebst2007,tupitsyn2010}. For the error model
we adopt, if the coupling constants in $\tilde{H}$ are homogeneous,
the spin system $\{\mu_{\bf m}\}$ does not behave as a spin glass.

We now discuss in more detail the critical behavior of the
boundary-field Ising model in some special situations and how that
behavior affects the surface code threshold.

\subsection{Homogeneous coupling}

Consider $\tilde{J}_{{\bf u}{\bf v}} = 0$ and $\tilde{h}_{{\bf m}{\bf
    n}} = h < 0$ and real, corresponding to a single-qubit relaxation
channel. In this case Eq. (\ref{H_eff-2}) is reduced to the
ferromagnetic square lattice Ising model with a boundary field. In
particular, ${\cal A}_{0}$ and ${\cal B}_{0}$ are determined by the
partition function of this model \cite{mccoy1967abraham1980}. It is
known that the free energy has two different terms: a boundary and a
bulk contribution. Although the boundary magnetization has a different
exponent than the bulk one, the critical temperature for the
ferromagnetic transition is defined by the bulk transition temperature
\cite{onsager}, $\left| h_{\rm critical} \right| = \ln \left( 1 +
\sqrt{2} \right)/2$. In the high-temperature paramagnetic phase,
$\left| h \right| < \left| h_{\rm critical} \right|$, the direction of
the boundary fields is irrelevant. Hence, in this fully $Z_{2}$
symmetric phase, we find that ${\cal B}_{0}\to0$ in the thermodynamic
limit. Below the critical temperature, $\left| h \right| > \left|
h_{\rm critical} \right|$, the boundary fields explicitly break the
$Z_{2}$ symmetry, leading to two distinct values for ${\cal B}_{0}$
when $\alpha_t = \alpha_b$ and $\alpha_t \neq \alpha_b$. In this
phase, ${\cal B}_{0} \neq 0$ in the thermodynamic limit and we find
${\cal F}_0 < 1$. The transition is exponentially sharp since the
boundary free energy is proportional to the number of sites at the
edge. For other syndromes, a very similar discussion can be made. In
the thermodynamic limit, the transition to a regime where the code can
correct happens simultaneously for all syndromes since the critical
point is entirely controlled by $\tilde{H}_{\rm eff}$.

\begin{figure}
\centering
\includegraphics[width=7cm]{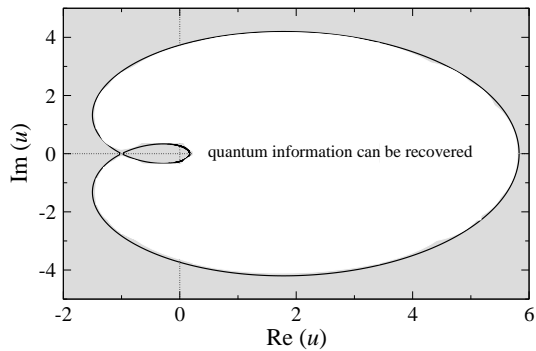}
\caption{Phase diagram of the Ising model with homogeneous complex
  coupling constant $h$ and its significance to the one-cycle QEC
  fidelity threshold. Here $u = e^{-4h}$. In the gray regions quantum
  information cannot be recovered.}
\label{fig:hr_hi_full}
\end{figure}

This analysis can be extended to complex $h$ \cite{matveev1995}. In
this more general situation, the transition point between the $Z_{2}$
symmetric phase and the broken symmetry phase is defined by the curve
$\sinh^{2} \left( 2h_{\rm critical} \right) = e^{-i\theta}$, where
$\theta \in [0,2\pi)$. See Fig. \ref{fig:hr_hi_full}. For a purely
  dynamical problem, when $h$ is imaginary (e.g., when a uniform
  external magnetic field acts on the physical qubits), $h_{\rm
    critical} = \pm i\pi/4$. In this case the critical point
  corresponds to $p_c=1/2$, which is consistent with the code
  providing infinite protection in the thermodynamic limit.

Now, consider $\tilde{h}_{{\bf m}{\bf n}} = 0$ and $\tilde{J}_{{\bf
    u}{\bf v}} = \tilde{J} < 0$. This case applies to qubits coupled
linearly to a gapless bosonic bath \cite{novais2013,jouzdani2013}. The
Hamiltonian in Eq. (\ref{H_eff-2}) can be broken into two independent
square-lattice Ising models with nearest-neighbor interactions. Hence,
the discussion from the previous paragraph can be immediately
applied. Note that $\tilde{J}$ is doubled with respect to its value
for the physical qubit interactions, i.e., $\tilde{J} = 2J$.

\subsection{Random coupling}

Let $\tilde{J}_{{\bf u}{\bf v}} = 0$ and $\tilde{h}_{{\bf m}{\bf n}}$
be real and random. The Harris criterion can not be applied since, for
the clean Ising model, the specific heat critical exponent vanishes
and the model is marginal to disorder \cite{lykke1998}. We therefore
discuss some specific cases. If $\tilde{h}_{{\bf m}{\bf n}}$ has the
same sign for all bonds, we expect bond disorder to be perturbative
and simply yield a transition temperature roughly given by the typical
value of $\tilde{h}_{{\bf m}{\bf n}}$. This can be put on firm grounds
by considering a simple toy model with two possible values for the
bond, $\tilde{h}_{{\bf m}{\bf n}} = h_{1}$ or $h_{2}$, with equal
probability \cite{cardy1997}. In this case the transition temperature
is given by $\left( e^{h_{1}} - 1 \right) \left( e^{h_{2}} - 1 \right)
= 2$.

An interesting situation arises when there is bond dilution, namely,
when some of the $\tilde{h}_{{\bf m}{\bf n}}$ are equal to zero. In
the case of a square lattice, the percolation threshold happens when
half of the bonds are missing. Thus, if at least half of the qubits do
not suffer the action of the local magnetic field $h_{{\bf i}}$, the
probability of having an infinite cluster tends to zero and the
$Z_{2}$ broken phase does not exist. The implication to QEC is that
quantum information can \emph{always} be recovered if at least half of
the qubits do not fail during a QEC cycle. Thus, only a severe random
event that causes most of the qubits to malfunction during the
computation will lead to a failure in the computation. This can be
relevant to the design of other planar codes as well (which can be
engineered to have high percolation thresholds). 

\begin{figure}[t]
\centering
\includegraphics[width=6cm]{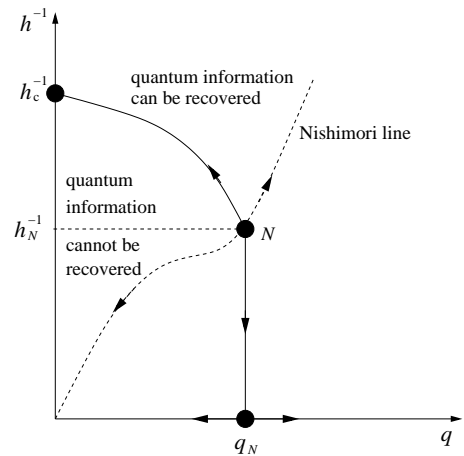}
\caption{Phase diagram of the Ising model with random bond sign and
  its significance to the one-cycle QEC fidelity threshold.}
\label{fig:nishimori-diag}
\end{figure}

The scenario dramatically changes when we allow bonds with different
sign \cite{honecker2001}. A Gaussian distribution is likely a
realistic assumption for this case \cite{mcmillian1984}, but most of
the physics can already be discussed using the toy-model bond
distribution $P\left( h,q \right) = q \delta \left( \tilde{h}_{{\bf
    m}{\bf n}}-h \right) + \left( 1 - q \right) \delta \left(
\tilde{h}_{{\bf m}{\bf n}} + h \right)$. There are three
renormalization group fixed points for this model in the $\left( h, q
\right)$ plane \cite{honecker2001} (see
Fig. \ref{fig:nishimori-diag}). On top of the well-known Nishimori
line, there is an unstable fixed point, $\left( h_{N},q_{N} \right)$,
that separates the $Z_{2}$ broken phase from the unbroken phase. For
$q < q_{N}$ and $h < h_{N}$ the physics is controlled by the stable
fixed point of the model $\left( h_c,0 \right)$, hence we fall back to
the discussion of the homogeneous model and get the usual
paramagnetic-ferromagnetic transition. Conversely, for $q \geq q_{N}$
and $h > h_{N}$ the transition is controlled by the fixed point with
$\left( \infty,q_{N} \right)$. Little is known about this fixed point,
but it is believed that it separates the ferromagnetic phase from a
spin glass phase at $h = \infty$. The existence of the $Z_{2}$ broken
phase is not in question, but the nature of the unbroken phase, where
we expect the fidelity to be higher, is not clear and needs further
investigation.

Finally, let us consider $\tilde{h}_{{\bf m}{\bf n}} = 0$ and
$\tilde{J}_{{\bf u}{\bf v}}$ real and random: All the discussion from
the previous paragraphs can be immediately transported to this case,
the main difference being that there are two independent
lattices. Hence, as a function of disorder or coupling strength, one
can have two different transition temperatures and thus a more
complicated threshold situation may arise.

\section{Numerical simulations}
\label{sec:numerics}

\begin{figure}[t]
\centering
\includegraphics[width=7cm]{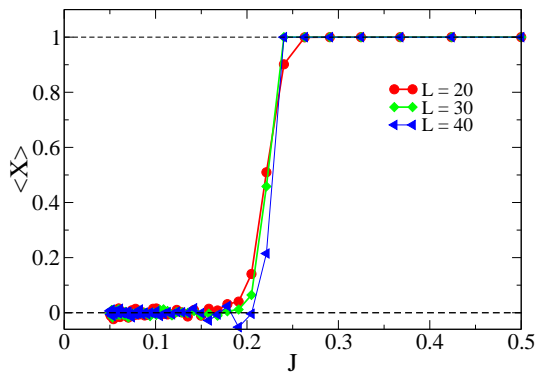}
\caption{(Color online) Ratio $\langle X \rangle =
  |\frac{\mathcal{B}}{\mathcal{A}}|$ evaluated through Monte Carlo
  sampling for three system sizes: $L=20$ ($761$ qubits, circles),
  $L=30$ ($1741$ qubits, diamonds), and $L=40$ ($3121$ qubits,
  triangles). The solid lines are guides to the eyes.}
\label{fig:X} 
\end{figure}

Some of the results described above were independently confirmed by
Monte Carlo simulations. Here, we present the case of a constant and
real $J_{{\bf i}{\bf j}} = J$ for nearest neighbors, $h = 0$, and
nonerror syndromes. To insert the constraint $A_{\diamondsuit} = 1$
into the Metropolis algorithm, an alternative representation of the
stabilizers was necessary. The stabilizer operator was rewritten as
the product of an even number $m$ of logical operators such that their
product is equivalent to the stabilizer, namely, $A_{\diamondsuit} =
\bar{Z}_{\Gamma_{1}} \dots \bar{Z}_{\Gamma_{2m}}$ (see
Ref. \cite{pejman_numerics} for details of this formulation and more
extensive numerical results). Working directly with the original spin
(i.e., physical qubit) variables, we used this representation to
numerically evaluate the amplitudes ${\cal A}_{0}$ and ${\cal B}_{0}$
and the ratio $\langle X \rangle = |{\cal B}_{0}| / |{\cal
  A}_{0}|$. Data for the case of nearest-neighbor interactions are
shown in Fig. \ref{fig:X}. Notice that the larger the lattice, the
sharper the transition becomes. The mapping onto the unconstrained
Ising model predicts that the critical coupling $J_c$ should be equal
to half of that for a regular two-dimensional Ising system, namely,
$J_c \approx 0.220$. This is in excellent agreement with the numerical
value of $J_c = 0.217$ obtained in our Monte Carlo simulations through
a finite-size scaling analysis.

\section{Conclusions}
\label{sec:conclusions}

In this paper, we have argued that a stabilizer code, such as the
surface code, when coupled to an environment, has an intrinsic
fidelity threshold. This threshold can be determined by focusing on
the nonerror syndrome sector after the physical qubits have evolved in
time under an effective action intermediated by the environment. For
other syndrome sectors, the decoding of the syndrome into a recovery
operation limits the error correction capabilities of the code. As a
result, the effective fidelity threshold should be smaller than the
intrinsic threshold. To show that a threshold exists nevertheless for
any syndrome, we provide a prescription where the decoding always work
in the infinite lattice size limit. When bit-flip errors are
considered, this prescription allows us to derive an expression for
the fidelity of the surface code in terms of two amplitudes that
involve expectation values of a constrained classical spin model.

For the case of an effective action involving at most nearest-neighbor
interactions, we map the spin model onto an unconstrained Ising model
with boundary fields. This mapping allows us to predict the exact
value of the fidelity threshold for several important practical
situations. We find that a fidelity threshold is almost always
present, for both homogeneous and inhomogeneous noise sources, but the
actual critical value of the coupling constant between qubit and
environment depends on particular features of the model. Thus, the
threshold is not universal. For the case of homogeneous coupling to
the environment, the analytical prediction for the threshold location
based on the exact mapping is confirmed by an independent Monte-Carlo
simulation.

When the effective interaction between physical qubits goes beyond
nearest neighbors, the mapping no longer works, but the threshold can
still be computed by direct numerical simulations of the constrained
spin model. A recent work by two of the authors \cite{pejman_numerics}
shows that the threshold decreases with increasing interaction range,
as originally predicted in Ref. \cite{novais2013}.

\acknowledgments

We are grateful to Robert Raussendorf for inspiring discussions. We
thank Jos\'e Hoyos Neto and Barbara Terhal for useful
conversations. E.N. was partially supported by INCT-IQ and
CNPq. P.J. and E.R.M. were supported in part by the Office of Naval
Research and the National Science Foundation under Grant
No. CCF-1117241. I.S.T. acknowledges support from IARPA and PITP.


\appendix

\section{Existence of an intrinsic threshold}
\label{sec:appendixA}

Let us consider the case of nonerror syndrome and assume that the
environment is reset to its ground state at the end of the QEC
cycle. The final state of the physical qubits is given by the state
vector
\begin{equation}
| \psi_{\rm QEC} \rangle_0 = {\cal P}_0\, U_{\rm eff} | \bar{\uparrow}
\rangle,
\end{equation}
where
\begin{equation}
| \bar{\uparrow} \rangle = \frac{1}{\sqrt{2^{N_{\diamond}}}}
\prod_{\lozenge} ( 1 + A_{\lozenge}) | F_z \rangle,
\end{equation}
$| F_z \rangle$ is the ferromagnetic $z$ state, and the projector of
positive stars and plaquettes is
\begin{equation}
{\cal P}_0 = \frac{1}{2^{N_{\lozenge}} 2^{N_{\square}}}
\prod_{\lozenge} ( 1 + A_{\lozenge}) \prod_{\square} ( 1 +
B_{\square}) .
\end{equation}
Since ``no errors'' were identified, there is no need to apply a
recovery operation: ${\cal S}^x_0 = 1$ (i.e., the identity operator).

Introducing the amplitudes
\begin{eqnarray}
\mathcal{A}_0 & = & \langle \bar{\uparrow} |\psi_{\rm QEC}\rangle_0
\nonumber \\ & = & \frac{1}{2^{N_{\lozenge}} 2^{N_{\square}}} \langle
\bar{\uparrow} | \prod_{\lozenge} ( 1 + A_{\lozenge}) \prod_{\square}
( 1 + B_{\square}) U_{\rm eff} | \bar{\uparrow} \rangle
\end{eqnarray}
and
\begin{eqnarray}
\mathcal{B}_0 & = & \langle \bar{\downarrow} |\psi_{\rm QEC}\rangle_0
\nonumber \\ & = & \frac{1}{2^{N_{\lozenge}} 2^{N_{\square}}} \langle
\bar{\uparrow} | \bar{X} \prod_{\lozenge} ( 1 + A_{\lozenge})
\prod_{\square} ( 1 + B_{\square}) U_{\rm eff} | \bar{\uparrow}
\rangle,
\end{eqnarray}
we can express the fidelity without any ambiguity as
\begin{equation}
\mathcal{F}_0 = \frac{| \mathcal{A}_0 |^2}{| \mathcal{A}_0 |^2 + |
  \mathcal{B}_0 |^2} .
\end{equation}

In order to evaluate the amplitudes, it is convenient to move to the
$x$ basis $\{ | s \rangle \}$. Since
\begin{equation}
  | F_z \rangle = \prod_{i = 1}^N \left( \frac{| \uparrow \rangle_{x,
      i} + | \downarrow \rangle_{x, i}}{\sqrt{2}} \right) =
  \frac{1}{\sqrt{2^N}} \sum_s | s \rangle,
\end{equation}
we arrive at
\begin{equation}
\mathcal{A}_0 = \frac{1}{2^{N_{\square}} 2^N} \sum_s \langle s |
\prod_{\lozenge} ( 1 + A_{\lozenge}) \prod_{\square} ( 1 +
B_{\square}) U_{\rm eff} | s \rangle
\end{equation}
and
\begin{equation}
\mathcal{B}_0 = \frac{1}{2^{N_{\square}} 2^N} \sum_s \langle s |
\bar{X} \prod_{\lozenge} ( 1 + A_{\lozenge}) \prod_{\square} ( 1 +
B_{\square}) U_{\rm eff} | s \rangle,
\end{equation}
where we used that $\frac{1}{2^{N_{\lozenge}}} \left[ \prod_{\lozenge}
  ( 1 + A_{\lozenge}) \right]^2 = \prod_{\lozenge} ( 1 +
A_{\lozenge})$. Substituting $U_{\rm eff} = e^{- H_{\rm eff}}$ and
recalling that, for bit-flip errors, $H_{\rm eff}$ is diagonal in the
$x$ basis, we can write
\begin{equation}
\label{eq:A0}
\mathcal{A}_0 = \frac{1}{2^{N_{\square}} 2^N} \sum_s e^{- E_s} \langle s
| \prod_{\lozenge} ( 1 + A_{\lozenge}) \prod_{\square} ( 1 +
B_{\square}) | s \rangle
\end{equation}
and
\begin{equation}
\label{eq:B0}
\mathcal{B}_0 = \frac{1}{2^{N_{\square}} 2^N} \sum_s e^{- E_s} \langle s
| \bar{X} \prod_{\lozenge} ( 1 + A_{\lozenge}) \prod_{\square} ( 1 +
B_{\square}) | s \rangle,
\end{equation}
where $H_{\rm eff} | s \rangle = E_s | s \rangle$.

At this point the sums over states $s$ contain all possible $x$ spin
configurations of the physical qubits. However, the projectors in
Eqs. (\ref{eq:A0}) and (\ref{eq:B0}) will restrict these
configurations. In order to proceed, we can explicitly write that any
state $s$ that satisfies the projectors can be written in one of the
two forms:
\begin{eqnarray}
  | s_1 \rangle & = & \prod_j B_j | F_x \rangle \\
  | s_2 \rangle & = & \bar{Z}_{\gamma}  \prod_j B_j | F_x \rangle, 
\end{eqnarray}
where $|F_x\rangle$ is the ferromagnetic state in the $x$ direction,
$\bar{Z}_{\gamma}$ is a logical $z$ following a path $\gamma$, and
$\prod_j B_j$ is a product over a set of plaquettes. In order for the
basis $\{ | s_1 \rangle, | s_2 \rangle \}$ to be complete, all
possible products of plaquettes have to be used when generating the
states $|s_1\rangle$. Notice that only one path $\gamma$ should be
used for generating the states $|s_2\rangle$; summing over more than
one path will render the basis overcomplete. In principle the choice
of $\gamma$ should not be important in the evaluation of the fidelity
(choosing $\gamma$ amounts to choosing a gauge; for each state
$|s_1\rangle$ there is a state $|s_2\rangle$ and different $\gamma$'s
just define different correspondences between those states).

Thus, the restricted sums can be explicitly written as
\begin{eqnarray}
\mathcal{A}_0 & = & \frac{1}{2^N} \left( \sum_{s_1} e^{- E_{s_1}}
\langle s_1 |s_1 \rangle + \sum_{s_2} e^{- E_{s_2}} \langle s_2 |s_2
\rangle \right) \nonumber \\ & = & \frac{1}{2^N} \left( \sum_{s_1}
e^{- E_{s_1}} + \sum_{s_2} e^{- E_{s_2}} \right)
\end{eqnarray}
and
\begin{eqnarray}
\mathcal{B}_0 & = & \frac{1}{2^N} \left( \sum_{s_1} e^{- E_{s_1}}
\langle s_1 | \bar{X} | s_1 \rangle + \sum_{s_2} e^{- E_{s_2}} \langle
s_2 | \bar{X} | s_2 \rangle \right) \nonumber \\ & = & \frac{1}{2^N}
\left( \sum_{s_1} e^{- E_{s_1}} - \sum_{s_2} e^{- E_{s_2}} \right) .
\end{eqnarray}
Notice that $| \mathcal{A}_0 | \geq | \mathcal{B}_0 |$ always.

Now consider the infinite-lattice limit. When the spin model of
Eq. (\ref{eq:Heff}) has a well-defined phase transition at a finite
temperature, in the ``high-temperature'' (disordered) phase, which
here corresponds to small coupling constant values, states of type
$|s_1\rangle$ and $|s_2\rangle$ have the same ``partition function'',
namely, $\sum_{s_1} e^{- E_{s_1}} = \sum_{s_2} e^{- E_{s_2}}$. As a
result, $\mathcal{B}_0 = 0$ and $\mathcal{F}_0 = 1$ for coupling
constant values below the critical point. In the ``low-temperature''
(ordered) phase, states $|s_2\rangle$ are energetically more costly
than states $|s_1\rangle$ \cite{novais2013}, resulting in $\left(
\sum_{s_1} e^{- E_{s_2}} \right) / \left( \sum_{s_2} e^{- E_{s_1}}
\right) \rightarrow 0$ in the thermodynamic limit. Therefore, for
coupling constant values above the critical point, $\mathcal{B}_0 =
\mathcal{A}_0$ and $\mathcal{F}_0 = 1/2$.




\begin{thebibliography}{35}

\bibitem{shor} P. W. Shor, Phys. Rev. A {\bf 52}, R2493 (1995).

\bibitem{gottesman} D. Gottesman, Ph.D. thesis, Caltech, 1997.

\bibitem{bravyi1998} S. B. Bravyi and A. Y. Kitaev,
  arXiv:quant-ph/9811052.

\bibitem{dennis2002} E. Dennis, A. Kitaev, A. Landahl, and
  J. Preskill, J. Math. Phys. \textbf{43}, 4452 (2002).

\bibitem{fowler2012b} A. G. Fowler, M. Mariantoni, J. M. Martinis, and
  A. N. Cleland, Phys. Rev. A \textbf{86}, 032324 (2012).

\bibitem{raussendorf} R. Raussendorf, J. Harrington, and K. Goyal,
  Ann. Phys. {\bf 321}, 2242 (2006); New J. Phys. {\bf 9}, 199 (2007);
  R. Raussendorf and J. Harrington, Phys.  Rev. Lett. \textbf{98},
  190504 (2007).

\bibitem{wang2011} D. S. Wang, A. G. Fowler, and L. C. L. Hollenberg,
  Phys. Rev. A {\bf 83}, 020302(R) (2011).

\bibitem{ghosh2012} J. Ghosh, A. G. Fowler, and M. R. Geller,
  Phys. Rev. A {\bf 86}, 062318 (2012).

\bibitem{fowler} A. G. Fowler, Phys. Rev. Lett. {\bf 109}, 180502
  (2012); A. G. Fowler, arXiv:1310.0863; A. G. Fowler and
  J. M. Martinis, Phys. Rev. A {\bf 89}, 032316 (2014).

\bibitem{stephens2013} A. M. Stephens, Phys. Rev. A {\bf 89}, 022321
  (2014).

\bibitem{klesse2005} R. Klesse and S. Frank, Phys. Rev. Lett. {\bf
  95}, 230503 (2005).

\bibitem{novais2007} E. Novais, E. R. Mucciolo, and H. U. Baranger,
  Phys. Rev. Lett. {\bf 98}, 040501 (2007); Phys. Rev. A {\bf 82},
  020303(R) (2010).

\bibitem{preskill} J. Preskill, Quant. Inf. Comput. {\bf 13}, 181
  (2013); H. K. Ng and J. Preskill, Phys. Rev. A {\bf 79}, 032318
  (2009).

\bibitem{kitaev} A. Yu. Kitaev, Ann. Phys. {\bf 303}, 2 (2003).

\bibitem{trebst2007} S. Trebst, P. Werner, M. Troyer, K. Shtengel, and
  C. Nayak, Phys. Rev. Lett. {\bf 98}, 070602 (2007).

\bibitem{tupitsyn2010} I. S. Tupitsyn, A. Kitaev, N. V. Prokof'ev, and
  P. C. E. Stamp, Phys. Rev. B {\bf 82}, 085114 (2010).

\bibitem{terhalRMP} B. Terhal, arXiv:1302.3428.

\bibitem{superconductor} J. Clarke and F. K. Wilhelm, Nature (London)
  {\bf 453}, 1031 (2008); D. P. DiVincenzo, Phys. Scr. {\bf T137},
  014020 (2009).

\bibitem{coldatoms} I. Bloch, J. Dalibard, and W. Zwerger, Rev. Mod.
  Phys. \textbf{80}, 885 (2008).

\bibitem{trappedions} D. Nigg {\it et al.}, Science {\bf 345}, 302
  (2014).

\bibitem{rydberg} M. Saffman, T. G. Walker, and K. M\o lmer,
  Rev. Mod. Phys. {\bf 82}, 2313 (2010).

\bibitem{semiconductor} M. A. Eriksson {\it et al}., Quantum
  Inf. Proc. {\bf 3}, 133 (2004); H. Bluhm {\it et al}.,
  Nat. Phys. {\bf 7}, 109 (2011).

\bibitem{obs1} The allowed $\Gamma_Z$ paths can be obtained from that
  shown in Fig. \ref{fig:surfacecode} by pulling it up or down at the
  sites on horizontal edges. Similarly, allowed $\Gamma_X$ paths are
  generated from that shown in Fig. \ref{fig:surfacecode} by pulling
  it to left or to the right at the sites on horizontal edges.

\bibitem{novais2013} E. Novais and E. R. Mucciolo, Phys. Rev. Lett.
  \textbf{110}, 010502 (2013).

\bibitem{jouzdani2013} P. Jouzdani, E. Novais, and E. R. Mucciolo,
  Phys. Rev. A \textbf{88}, 012336 (2013).


\bibitem{poulin} G. Duclos-Cianci and D. Poulin, Phys. Rev. Lett. {\bf
  104}, 050504 (2010).

\bibitem{bravyi} S. Bravyi, M. Suchara, and A. Vargo,
  arXiv:1405:4883.

\bibitem{ferris} A. J. Ferris and D. Poulin, Phys. Rev. Lett. {\bf
  113}, 030501 (2014).

\bibitem{stamp} P. C. E. Stamp and I. S. Tupitsyn, Chem. Phys. {\bf
  296}, 281 (2004); A. Morello, P. C. E. Stamp, and I. S. Tupitsyn,
  Phys. Rev. Lett. {\bf 97}, 207206 (2006).



\bibitem{polyakov} A. M. Polyakov, {\it Gauge Fields and Strings}
  (Hardwood Academic Publishers, London, 1987).

\bibitem{mccoy1967abraham1980} B. M. McCoy and T. T. Wu,
  Phys. Rev. \textbf{162}, 436 (1967); D. B. Abraham,
  Phys. Rev. Lett. {\bf 44} 1165 (1980); H. Au-Yang and M. E. Fisher,
  Phys. Rev. B {\bf 21} 3956 (1980); A. Maciolek and J.Stecki, {\it
    ibid}.  {\bf 54} 1128 (1996).

\bibitem{onsager} L. Onsager, Phys. Rev. {\bf 65}, 117 (1944).

\bibitem{matveev1995} V. Matveev and R. Shrock, J. Phys. A:
  Math. Gen. {\bf 28} 1557 (1995)

\bibitem{lykke1998} J. L. Jacobsen and J. Cardy, Nucl. Phys. B {\bf
  515}, 701 (1998).

\bibitem{cardy1997} J. Cardy and J. L. Jacobsen,
  Phys. Rev. Lett. {\bf 79}, 4063 (1997).

\bibitem{honecker2001} A. Honecker, J. L. Jacobsen, M. Picco, and
  P. Pujol, in Proceedings of the NATO Advanced Research Workshop on
  Statistical Field Theories, edited by A. Cappelli and G. Mussardo
  (Kluwer Academic, Dordrecht, 2002); arXiv/cond-mat/0112069.

\bibitem{mcmillian1984} W. L. McMillan, Phys. Rev. B {\bf 29}, 4026
  (1984).

\bibitem{pejman_numerics} P. Jouzdani and E. R. Mucciolo, Phys. Rev. A
  {\bf 90}, 012315 (2014).

\end{thebibliography}
\end{document}